\documentclass[journal]{IEEEtran}

\usepackage{amssymb}
\usepackage{amsmath} 
\usepackage{mathtools}
\usepackage{booktabs}
\usepackage{graphicx}
\usepackage{color}

\DeclareMathOperator{\rect}{rect}

\ifCLASSINFOpdf
\else
\fi

\begin{document}

\title{Information Rate in Ultra-Wideband Optical Fiber Communication Systems Accounting for High-Order Dispersion}

\author{Nikita~A.~Shevchenko, Tianhua~Xu~\IEEEmembership{Member,~IEEE}, Cenqin~Jin, Domani\c{c}~Lavery~\IEEEmembership{Member,~IEEE}, Robert~I.~Killey~\IEEEmembership{Senior~Member,~IEEE} and Polina~Bayvel~\IEEEmembership{Fellow,~IEEE,~OSA} 
\thanks{This work is supported by UK~EPSRC Program TRANSNET and EU~Horizon~2020~RISE~Grant~778305. $(Corresponding~author: Tianhua~Xu)$}
\thanks{N.~A.~Shevchenko is with the Department of Physics, King's College London, London WC2R 2LS, UK (e-mail: mykyta.shevchenko@kcl.ac.uk).}
\thanks{T.~Xu and C.~Jin are with the School of Engineering, University of Warwick, Coventry CV4 7AL, UK and T. Xu is also with the Optical Networks Group, University College London, London WC1E 7JE, UK (e-mail: tianhua.xu@ieee.org).}
\thanks{D.~Lavery, R.~I.~Killey and P.~Bayvel are with the Optical Networks Group, University College London, London WC1E 7JE, UK (e-mail: p.~bayvel@ucl.ac.uk).}
}


\maketitle

\begin{abstract}
The effect of Kerr-induced optical fiber nonlinearities in $C-$band ($\sim$5~THz) EDFA and $C+L-$band ($\sim$12.5~THz) Raman-amplified optical communication systems has been studied considering the impact of third-order fiber dispersion. The performance of digital nonlinearity compensation with single-channel and 250-GHz bandwidth in both EDFA and Raman amplified systems has been investigated, respectively. The achievable information rates (AIRs) and optimum code rates in each individual transmission channel have been evaluated for the DP-64QAM, the DP-256QAM and the DP-1024QAM modulation formats, both with and without the use of the probabilistic shaping technique. It is found that, for all considered modulation formats, the signal-to-noise ratios, AIRs and code rates exhibit significantly asymmetric behavior about the central channel due to the presence of the third-order dispersion. This provides a new insight that the forward error correction schemes have to be optimized asymmetrically, on a per-channel basis, to maximize the overall throughput.
\end{abstract}


\IEEEpeerreviewmaketitle

\section{Introduction}

\IEEEPARstart{O}{ptical} fiber communications have achieved unprecedented growth and success over the past three decades and more than 95\% of digital data traffic is currently carried over optical fiber networks, which form a substantial part of the national and international communication infrastructure \cite{Imran2018}. The use of lumped erbium-doped amplification (EDFA) and distributed Raman amplifiers (DRA) negated the need for electronic re-generators and enabled dense wavelength-division-multiplexing (WDM) transmission, although the success and performance of these amplifier technologies are seen as limiting the usable fiber bandwidth up to approximately $\sim$5~THz for EDFA systems and $\sim$10--15~THz for Raman systems, respectively \cite{Agrawal2010,Keiser2010}. Apart from the bandwidth limitations, the achievable information rates (AIRs) of optical communication systems become inherently constrained by the launched power thresholds, as a consequence of nonlinear power-dependent ``noise-like'' distortions. These distortions occur due to the presence of the optical Kerr effect in fibers, which usually manifests itself as self-phase modulation (SPM), cross-phase modulation (XPM) and four-wave mixing (FWM) \cite{Mitra2001,Essiambre2012}.

Research on AIRs has been performed for $\sim$5~THz EDFA \cite{Poggiolini2014} and  for 4.3~THz DRA \cite{Bosco2011} optical transmission systems considering electronic dispersion compensation (EDC) only. In these works, the nonlinear distortion of the central wavelength channel has been regarded as indicative of the nonlinear distortions for all channels over the entire optical bandwidth when analyzing both the channel properties and the AIRs \cite{Curri2013}, although those nonlinear interference processes should be different for each channel due to different inter-channel nonlinear mixing. In our previous work, the behavior of Kerr-induced nonlinearities and the AIRs for each individual channel have been evaluated for the 4.3~THz EDFA and the 12.5~THz ($\sim$100~nm) Raman amplified optical communication systems accounting only for the second-order dispersion effect \cite{Nikshev2016, Semrau2017}. The influence of dispersion slope on the FWM efficiency has been recently explored experimentally \cite{ali2018}, however, in all aforementioned works, the modeling of the third-order dispersion was omitted. Under this simplifying assumption, the nonlinear distortions across the whole amplified bandwidth appeared to be ideally symmetric around the central channel; and the central channel exhibits the worst case among all WDM channels due to its stronger inter-channel interference. As a result, it was concluded that the forward error correction (FEC) schemes and their code rates have to be optimized accordingly in a symmetric manner about the central wavelength.

In this paper, built on our previous works, the nonlinear behaviors, the AIRs and the code rates in $C-$band ($\sim$5~THz) EDFA and $C+L-$band ($\sim$12.5~THz) DRA amplified  optical fiber communication systems are investigated, where the impact of both second-order and third-order dispersion are taken into account. These estimations have been separately carried out for each individual channel in Nyquist-spaced WDM transmission systems using multiple modulation formats including dual-polarization 64-ary quadrature amplitude modulation (DP-64QAM), DP-256QAM and DP-1024QAM.  The performance of computationally feasible digital nonlinearity compensation (NLC) schemes including single-channel NLC and 250-GHz NLC, which is the widest digital NLC bandwidth experimentally investigated \cite{Kai2017}, is examined for the 157-channel$\times$32~GHz ($\sim$5~THz) EDFA transmission systems and the 391-channel$\times$32~GHz ($\sim$12.5~THz) DRA-amplified transmission systems, where the application of the probabilistic shaping technique using discrete Maxwell-Boltzmann distribution has also been analyzed.

\section{Theoretical Model}

\subsection{Fiber nonlinear distortions model}

The performance of each individual channel, $k$, in a dispersion-unmanaged WDM transmission system, which is affected by both ASE noise and the optical Kerr effect in a fiber, is described by the so-called effective signal-to-noise ratio (SNR) \cite{Johan2013,shev2017}, which can be expressed as follows in the cases of EDC and NLC
\begin{equation} \label{snr_eff}
\begin{split}
&\mathrm{SNR}_{\,k} 
\triangleq \frac{P_{k}}{\sigma_{{\mathrm{eff},\,k}}^{2}}, \\ 
&\sigma_{{\mathrm{eff},\,k}}^{2}\approx N_{\mathrm{s}}\sigma_{\mathrm{ASE}}^{2} +\Delta \eta_{k} \, P_{k}^{3}, \\
&\Delta \eta_{k} \triangleq \eta_{k}(N_{\mathrm{s}},B) - \eta_{k}(N_{\mathrm{s}},B_{\,\mathrm{NLC}}),
\end{split}
\end{equation}
where $k$ is the WDM channel index, $N_{\mathrm{s}}$ is the total number of fiber spans in a link, $\sigma_{\mathrm{ASE}}^{2}$ represents the ASE noise power arising from the optical amplification process, and $B$ and $B_{\,\mathrm{NLC}}$ are the total WDM transmitted bandwidth and the NLC bandwidth, respectively. Note that the EDC case corresponds to $\eta_{k}(N_{\mathrm{s}},B_{\,\mathrm{NLC}})=0$ in Eq.~\eqref{snr_eff}. 

\begin{table}[t]
\caption{System parameter values}
\begin{center}
\begin{tabular}{l|c|c}
\toprule
Parameters & Values & Units \\
\midrule
\midrule
Carrier wavelength     & 1550 & nm \\
Symbol rate            & 32   & GBaud \\ 
Channel spacing        & 32   & GHz\\ 
EDFA bandwidth         & 40   & nm \\ 
DRA bandwidth          & 100  & nm\\ 
EDFA~NF                & 4.5  & dB\\ 
Signal attenuation     & 0.2  &  dB~km\textsuperscript{-1}  \\ 
Raman pump attenuation & 0.25  & dB~km\textsuperscript{-1}  \\ 
Fiber dispersion       & 17 &  ps~nm\textsuperscript{-1}~km\textsuperscript{-1} \\ 
\multicolumn{1}{c|}{$\beta_{2}$} & -- 21.67 &  ps\textsuperscript{2}~km\textsuperscript{-1} \\ 
Fiber dispersion slope & 0.067 & ps~nm\textsuperscript{-2}~km\textsuperscript{-1}\\ 
\multicolumn{1}{c|}{$\beta_{3}$} & 0.145 & ps\textsuperscript{3}~km\textsuperscript{-1}\\
Fiber nonlinearity     & 1.2  & W\textsuperscript{-1}~km\textsuperscript{-1}\\ 
Transmission distance  & 25$\times$80  & km \\ 
\bottomrule
\end{tabular}
\label{tab1}
\end{center}\vspace*{-18pt}
\end{table}

We assume that the optical gain $G=e^{\alpha L_{s}}$, with $\alpha$ being the fiber attenuation  parameter, is precisely equal to the single span loss, and also remains constant within a given channel spacing $\Delta f$. Therefore, the total ASE noise power per fiber span in the case of dual-polarization is given by the well-known expression
\begin{equation}\label{ase_power1}
\sigma_{\mathrm{ASE}}^{2} = 2  \left(e^{\alpha L_{s}}-1\right) n_{\mathrm{sp}}  \, hf_{0} \cdot  \Delta f,
\end{equation}
where $hf_{0}$ is the average photon energy, $f_{0}$ is the optical carrier frequency, $h$ is Planck's constant, and $n_{\mathrm{sp}}$ is the spontaneous-emission factor \cite{Agrawal2010}, which is related to the EDFA noise figure (NF) as $\mathrm{NF} \approx  2n_{\mathrm{sp}} $ assuming $G\gtrsim10$~dB.

The total variance of the ASE noise per fiber span over the channel due to optical counter-pumped fiber Raman amplification is defined as follows
\begin{equation} \label{DRA}
\sigma_{\mathrm{ASE}}^{2} = 2 \left(\kappa_{T}+1\right) \mathcal{N}_{\mathrm{phot}}  \, hf_{0} \cdot \Delta f ,
\end{equation}
where $\kappa_{T}$ is the temperature dependent phonon occupancy factor \cite[Eq.~(7)]{bromage}, and $\mathcal{N}_{\mathrm{phot}}$ denotes the number of spontaneously emitted photons, which is given in \cite[Eq.~(6)]{chinn}.

\begin{figure}[t]
\centering
\includegraphics[scale=1.1]{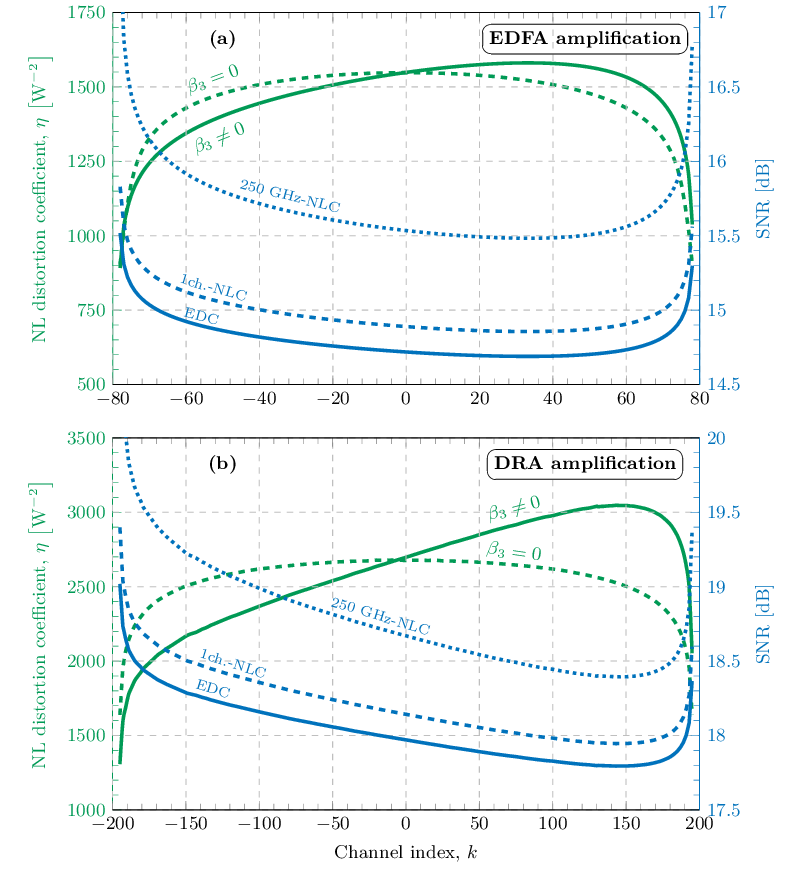}
\caption{Nonlinear distortion coefficient $\eta$ and SNR (at optimum launched power) for a fixed 2000~km (25$\times$80~km) transmission distance against WDM channel index. (a): lumped EDFA-amplified $C-$band ($\sim$~5~THz) system and (b): distributed Raman-amplified $\left(C+L\right)-$band ($\sim$~12.5~THz) system.}
\label{fig1}
\end{figure}


\begin{figure}[t]
\centering
\includegraphics[scale=1.1]{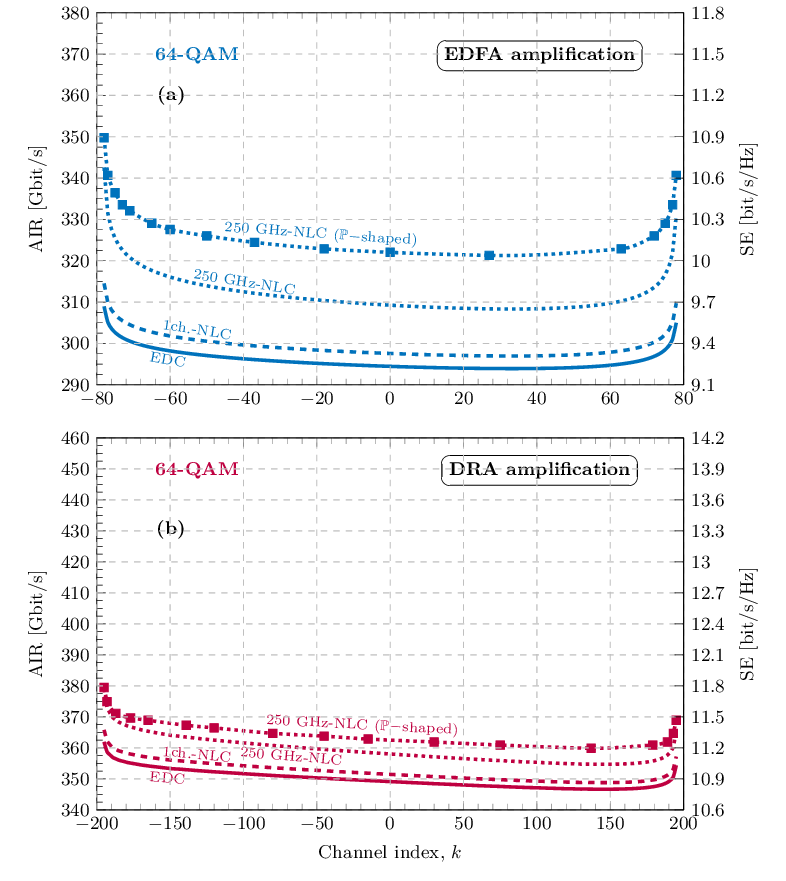}
\caption{AIRs (and SEs) for each individual channels for DP-64QAM optical communication systems. (a) EDFA system (b) Raman amplified system. $\mathbb{P}-$shaped: probabilistically shaped.}
\label{fig2}
\end{figure}

For dual-polarization Nyquist-spaced WDM transmission systems, which fulfill the Nyquist criterion, that is, having a rectangular spectra of width $\Delta f$ exactly equal to the symbol rate $R_{S}$, the nonlinear distortion coefficient $\eta$, which quantifies the impact of nonlinear distortions due to the FWM process, is given by \cite[Eq.~(4)]{Nikshev2016}
\begin{align} \label{eta_k}
\eta_{k}(N_{\mathrm{s}},B) 
= \frac{1}{\Delta f} \intop_{(2k-1) \,\Delta f/2}^{(2k+1)\, \Delta f/2}  \! \! \! \! \!  \! \! \!  df \, S\left(f; \,N_{\mathrm{s}},B\right) \, \rect \! \left(\frac{f}{\Delta f}\right),
\end{align}
where $\rect\left(x\right)$ denotes the rectangular function, and the Nyquist WDM channel index $k$ is given by the following set $k \triangleq \left\{ -\left(N_{\mathrm{ch}}-1\right)/2,\dots,\left(N_{\mathrm{ch}}-1\right)/2 \right\}$ with $N_{\mathrm{ch}}$ being the total number of WDM channels. Equation~\eqref{eta_k} implies the filtering of the ``noise-like" nonlinear distortion power spectral density (PSD) in the coherent receiver using a matched filter with a rectangular baseband transfer function. The PSD $S\left(f; \,N_{\mathrm{s}},B\right)$ of the nonlinear distortion is given by a slightly modified \cite[Eq.~(1)]{pogg2012}; it yields 
\begin{align}
\nonumber
S\left(f; \,N_{\mathrm{s}},B\right) 
&\approx \frac{16 \, \gamma^{\,2}}{27 R_{S}^2} \! \! \intop_{-B/2}^{\quad B/2} \! \! \! \intop_{-B/2}^{\quad B/2} \! \! df_{1} \, df_{2} \,  \left|\,\mu\left(f,f_{1},f_{2};\,N_{\mathrm{s}} \right)\,\right|^{\,2}  \\ 
&\cdot  \, \rect \! \left(\frac{f_{1} + f_{2} - f}{B}\right),
\end{align}
where $\gamma$ is the fiber Kerr nonlinearity parameter, $R_{S}$ is the symbol rate, $B\triangleq N_{\mathrm{ch}} \cdot \Delta f$ is the total transmitted bandwidth, and the kernel function can be factorised as follows
\begin{equation}
    \mu\left(f,f_{1},f_{2};\,N_{\mathrm{s}} \right) = \rho\left(f,f_{1},f_{2}\right) \cdot \phi\left(f,f_{1},f_{2}; \,N_{\mathrm{s}}\right),
\end{equation}
with the so-called phased-array factor $\phi\left(f,f_{1},f_{2};\,N_{\mathrm{s}}\right)$ being responsible for a distance evolution of the FWM nonlinear interaction process over the multi-span transmission system and being expressed as
\begin{equation}
\phi\left(f,f_{1},f_{2};\,N_{\mathrm{s}}\right) = \sum_{m=1}^{N_{\mathrm{s}}} e^{\,\imath \,\Delta \beta \left(f,f_{1},f_{2}\right)   \, \cdot \,  L_{\mathrm{s}} \, \left(m-1\right)},
\end{equation}
where $\imath \triangleq \sqrt{-1}$ is the imaginary unit, and $L_{\mathrm{s}}$ denotes the fiber span length. The FWM efficiently factor $\rho\left(f,f_{1},f_{2}\right)$ yields
\begin{equation} \label{fwm_factor}
\rho\left(f,f_{1},f_{2}\right) = \intop_{0}^{\: L_{\mathrm{s}}}  dz \,  e^{\,\imath\,\Delta \beta \left(f,f_{1},f_{2}\right) \cdot z} \, P(z) \,,
\end{equation}
where $L_{\mathrm{s}}$ is the fiber span length, $P(z)$ defines the signal power profile along a fiber span, which depends on the applied optical amplification scheme. For lumped EDFA case, the signal power profile $P(z)$ exponentially decays with distance, $P(z)=e^{-\alpha z}$. For the backward-pumped DRA case, the signal power profile $P(z)$ can be defined as a solution of a system of two coupled ordinary differential equations, which govern the Raman process for a single co-polarized pump wavelength and signal wavelength traveling in the backward direction \cite[Eqs.~(1-2)]{pelouch}.


\begin{figure}[t]
\centering
\includegraphics[scale=1.1]{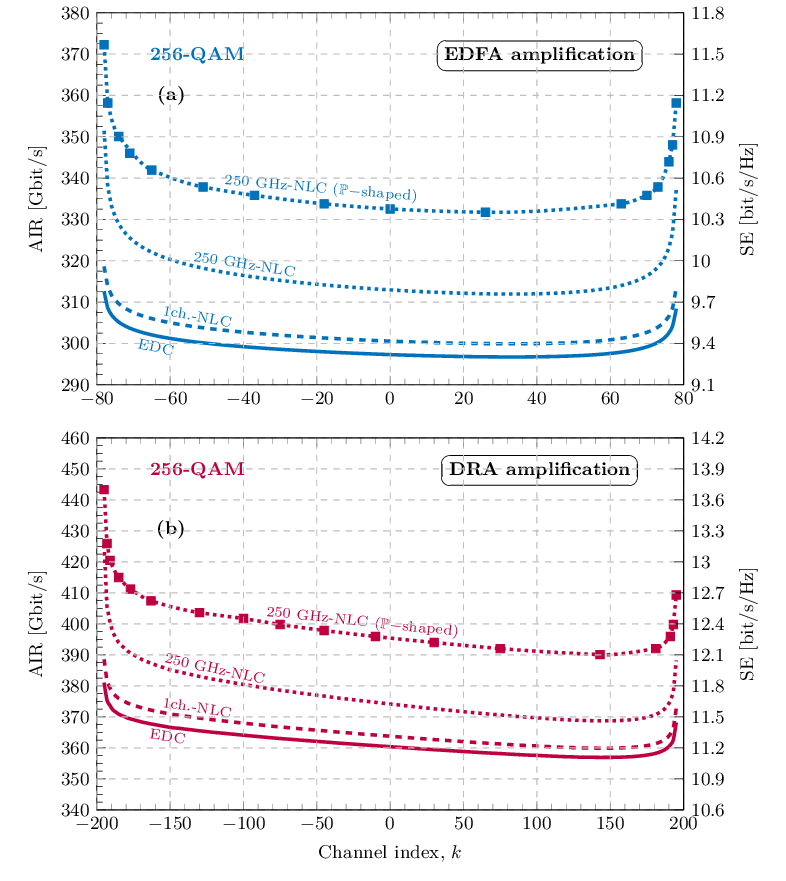}
\caption{AIRs (and SEs) for each individual channels for DP-256QAM optical communication systems. (a) EDFA system (b) Raman amplified system. $\mathbb{P}-$shaped: probabilistically shaped.}
\label{fig3}
\end{figure}


The FWM phase-mismatch $\Delta \beta$ can be approximated by including the third-order dispersion as follows \cite[Eq.~(5)]{zeiler1996}
\begin{align}
\nonumber
\Delta \beta  \left(f,f_{1},f_{2}\right) 
&=
\beta\left(f_{1} + f_{2} - f\right) \\ \nonumber
&- \beta \left(f_{1}\right) - \beta\left(f_{2}\right) + \beta\left(f\right) \\ 
\nonumber
&\approx 4 \pi^{2} \left[ \, \beta_{2} + \pi \left(f_{1} + f_{2} \right) \beta_{3} \, \right] \\
&\cdot \left(f_{1}-f\right)\left(f_{2}-f\right)\,,
\end{align} 
where $\beta \left(\cdot\right)$ denotes the propagation constant as a function of frequency, $\beta_{2}$ and $\beta_{3}$ represent the second-order and the third-order dispersion coefficients in the Taylor expansion, respectively \cite{Poggiolini2014,ali2018}. The impact of third-order dispersion on the distortion PSD of a 21-channel transmission system has been studied in \cite{Johan2013}, here we will investigate its impact in ultra-wideband transmission schemes and AIRs and code rates for each individual channel will be discussed.

\subsection{Estimation of achievable information rates}

The achievable information rate (AIR) for each individual channel can be estimated as in \cite{Semrau2017}
\begin{equation}
\mathrm{AIR} \triangleq 2 \, R_{S} \cdot I_{XY}.
\end{equation}
The ideal code rate $R^{\ast} \in \left[0,1\right]$ in the FEC scheme represents the maximum proportion of useful information in the coded bit sequence:
\begin{equation}
R^{\ast} \triangleq \frac{I_{X,Y}}{\log_{2} \left|\mathcal{X}\right|} \,,
\end{equation}
where $I_{X,Y}$ denotes the soft-decision mutual information (MI) between the sequence of output symbols $Y$ and input $X$ symbols. The MI is an important figure of merit as it yields a maximum information rate of a coded modulation scheme for which an arbitrarily small post-FEC BER can be achieved assuming that the input signal constellation drawn for an alphabet of cardinality $M\triangleq \left|\mathcal{X}\right|$ with a fixed PMF $p_{X}(x_{i})$, as well as a fixed channel law given by the conditional PDF $p_{Y|X}(y|x)$.

\begin{figure}[t]
\centering
\includegraphics[scale=1.1]{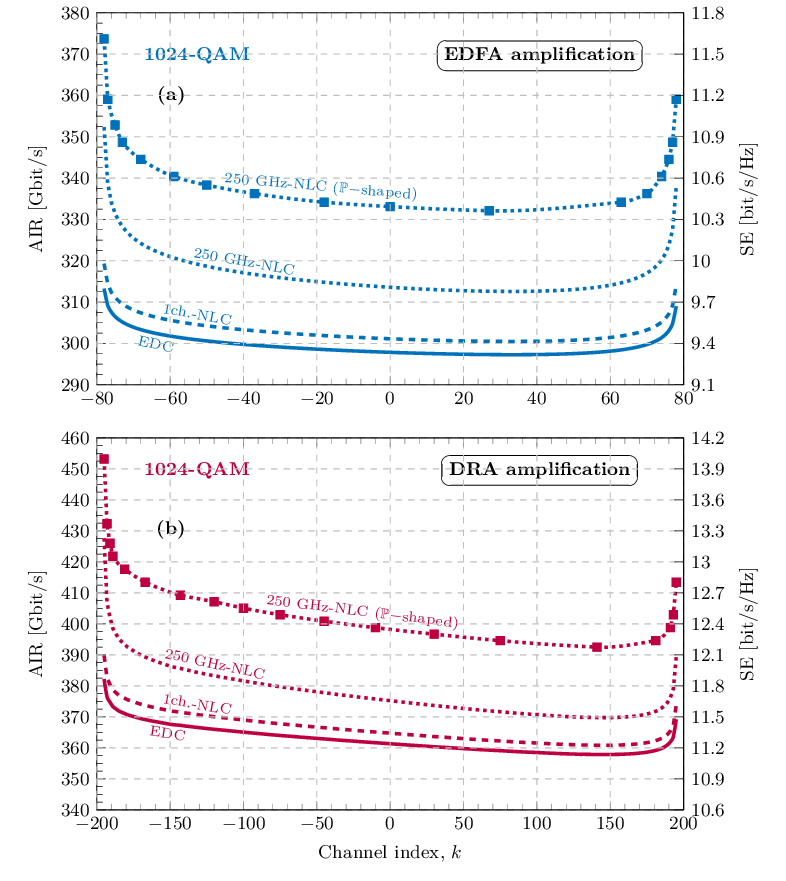}
\caption{AIRs (and SEs) for each individual channels for DP-1024QAM optical communication systems. (a) EDFA system (b) Raman amplified system. $\mathbb{P}-$shaped: probabilistically shaped.}
\label{fig4}
\end{figure}

Since the \emph{true} channel law for the nonlinear optical fiber channel is still subject to research, in this study the Gaussian channel law is applied by keeping consistency with the optical channel perturbative GN-model key assumption. Hence, owing to the information-theoretic AWGN channel assumption, the conditional PDF $p_{Y|X}(y|x)$ is set to be Gaussian, i.e.,
\begin{equation}
p_{Y|X}(y|x) = \frac{1}{\pi\sigma_{\mathrm{eff}}^{2}} \, \exp \left(-\frac{\left|\,y-x\,\right|^{2}}{\sigma_{\mathrm{eff}}^{2}}\right),
\label{gauss_ch_law}
\end{equation}
where the effective noise variance $\sigma_{\mathrm{eff}}^{2} = \mathbb{E}\left[\,\left|\,Y-X\right|^{2}\,\right]$ is assumed to be approximately equal to the effective noise variance in Eq.~\eqref{snr_eff}, and $\mathbb{E}\left[\cdot\right]$ denotes the mathematical expectation. It is worth mentioning that the estimated MI is an upper bound on the achievable rate under a Gaussian channel assumption, and it is not necessarily the greatest possible achievable rate for an arbitrarily complex receiver.

For a memoryless channel, the MI in bits per symbol for a fixed QAM input constellation is defined as
\begin{align} \label{MI_int} \nonumber
I_{X,Y} 
&= \sum_{i=1}^{M} \, \intop_{\mathbb{C}} \, dy  \, p_{Y|X}(y|x_{i}) \, p_{X}(x_{i}) \\ 
&\cdot \log_{2} \frac{p_{Y|X}(y|x_{i})}{\sum_{l=1}^{M} p_{Y|X}(y|x_{l}) \,  p_{X}(x_{l}) }\,,
\end{align}
where $\mathbb{C}$ represents the fixed set of complex numbers. The PMF $p_{X}(x_{i})\triangleq \mathbb{P} \left[X=x_{i}\right]$ in Eq.~\eqref{MI_int} of signal constellation points $\mathcal{X} \triangleq \left\{x_{1} \dots x_{M}\right\} \subseteq \mathbb{C}^{M}$ can be either uniform or with a shaped probability ($\mathbb{P}-$shaped).

In order to find the capacity-approaching probabilistically-shaped input QAM constellation, one needs to solve 2-D optimisation problem. However, due to the symmetry of the 2-D square QAM signalling, and the symmetry of typically used binary reflected Gray-coding, 2-D optimisation problem can be decomposed into a product of two constituent 1-D optimisation signal constellations of $\sqrt{M}$ points. It has been proven that an input QAM signal constellation probability shaping given by the discrete Maxwell-Boltzmann distribution is an optimal for AWGN channel \cite{Kschischang1993}. Theoretically, this yields to the well-known ultimate shaping gain of up to 1.53 dB \cite{forney1984}. Thus, the PMF in Eq.~\eqref{MI_int} is given by the family of expressions
\begin{equation} 
p_{X}(x_{i})  = 
    \begin{dcases} 
    \quad \quad  \quad \,\, 1/M, \quad  \quad  \quad  \quad \quad  \, \left(\text{equiprobable}\right) \\
    \frac{\exp\left( - \, \zeta \left|\,x_{i}\,\right|^{2} \right)}{\sum_{j=1}^{M}{\exp\left( - \, \zeta \left|\,x_{j}\,\right|^{2} \right)}}, \quad  \left(\mathbb{P}-\text{shaped}\right) 
    \end{dcases}
\end{equation}
where the scaling parameter $\zeta$ is chosen to maximize the MI, and $x_{i} \in \mathbb{C}$.

In particular, for equally likely input QAM constellation points, the MI in \eqref{MI_int} can be efficiently approximated by the following ready-to-use closed-form expression given by the Gauss-Hermite quadrature 
\begin{align} \label{mi_gh_approx_r2u}
\nonumber
I_{XY} 
&\approx \log_{2}M - \frac{1}{M\pi} \sum_{i=1}^{M} \sum_{k=1}^{L} w_{k} \sum_{l=1}^{L} w_{l} \\
&\cdot\log_{2} \sum_{j=1}^{M} \exp\left(-\,\frac{\left|\,d_{ij}\,\right|^{2}+2\,\sigma_{\mathrm{eff}}\,\Re\left[\left(\xi_{k}+\imath \,\xi_{l}\right)\,d_{ij} \right]}{\sigma_{\mathrm{eff}}^{2}}\right).
\end{align}
where $d_{ij}$ denotes the Euclidean distance between signal constellation symbols $x_{i}$ and $x_{j}$, $\xi_{k}$ and $\xi_{l}$ are the $k-$ and $l-$root of the $L-$order Hermite polynomial $H_{L}(x)$ with the associated weights $w_{k}$ and $w_{l}$, respectively.

\section{Results and Analyses}

\begin{figure}[t]
\centering
\includegraphics[scale=1.1]{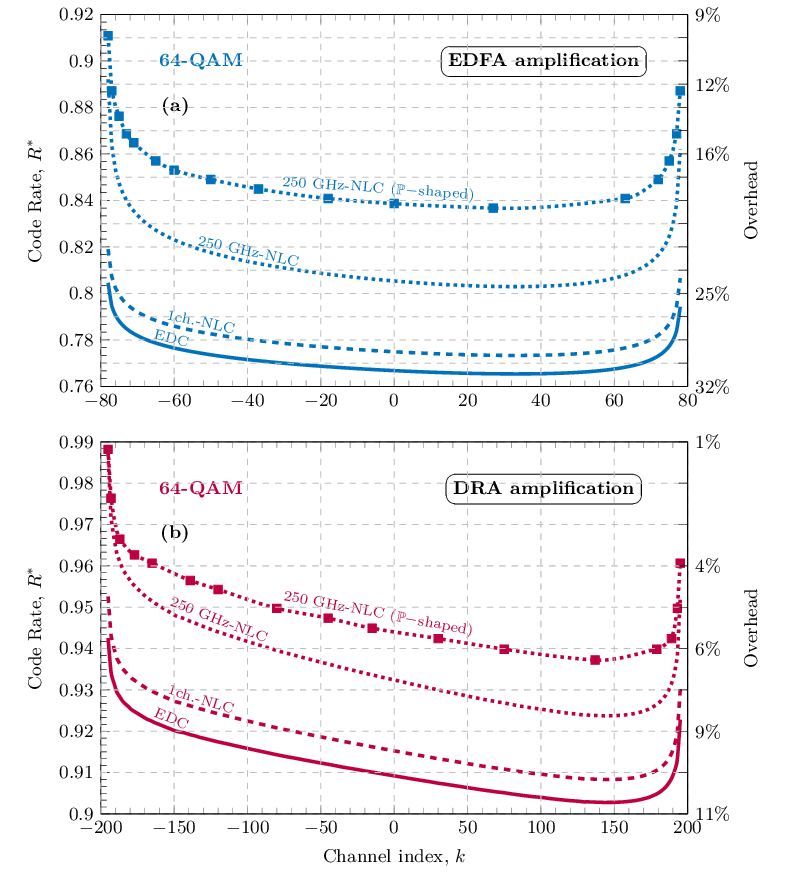}
\caption{Code rates (and overheads) for each individual channels for DP-64QAM optical communication systems. (a) EDFA system (b) Raman amplified system. $\mathbb{P}-$shaped: probabilistically shaped.}
\label{fig5}
\end{figure}

In this section, we have examined the backward-pumped geometry of the DRA scheme neglecting the Raman pump depletion effect as well as assuming the independence of the Raman gain on the laser wavelength. The optical lumped EDFA-amplified scheme was also analyzed. The nonlinear distortion model has been extended accounting for the effect of third-order dispersion (dispersion slope), which is captured by the term $\beta_{3}$. To investigate the differences in channel performance due to the impact of dispersion slope, the estimations have been carried out separately for each sub-channel in the ideal Nyquist-spaced WDM transmission system. Within the framework of the first-order pertubative analysis, the nonlinear distortion coefficients in Eq.~\eqref{eta_k} were numerically computed separately for each WDM channel via a quasi-Monte Carlo integration method \cite{Caflisch}, since the rate of convergence of the reference triple integral gradually decreases with increasing transmitted signal bandwidth. The transmission distance is fixed as 2000~km (25$\times$80~km). The EDC, the single-channel NLC and a practically feasible 250-GHz NLC, which was a reported record for the maximum possible signal processing bandwidth \cite{Kai2017} have been applied in the 157-channel$\times$32~GHz EDFA-amplified transmission system and 391-channel$\times$32~GHz DRA-amplified transmission system. The detailed parameters of the optical communication systems are summarized in Table~\ref{tab1}.

\begin{figure}[t]
\centering
\includegraphics[scale=1.1]{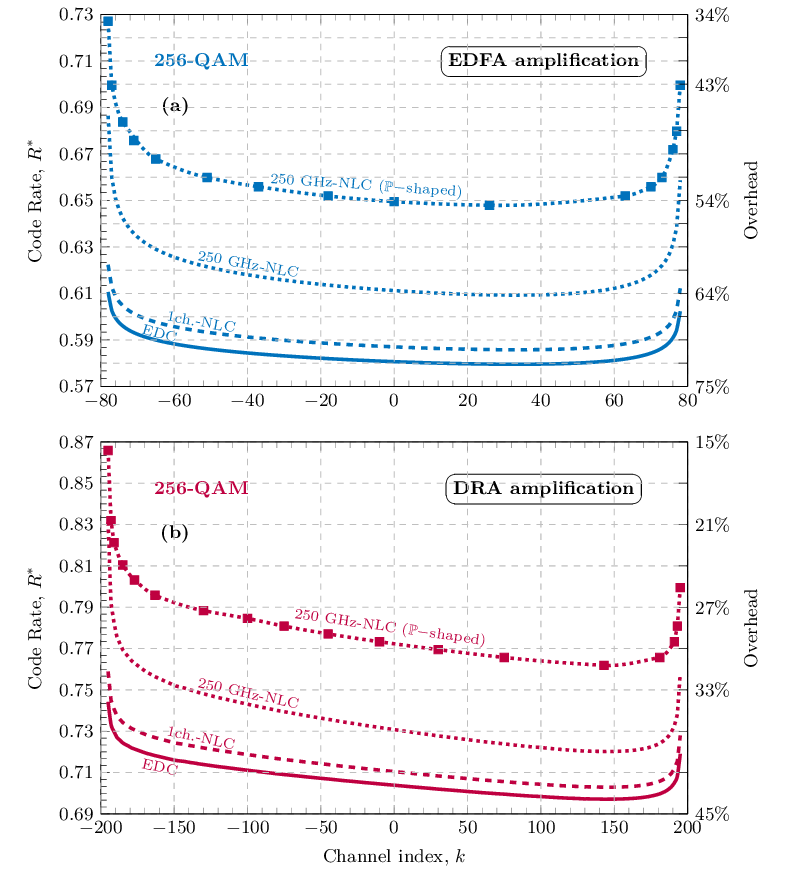}
\caption{Code rates (and overheads) for each individual channels for DP-256QAM optical communication systems. (a) EDFA system (b) Raman amplified system. $\mathbb{P}-$shaped: probabilistically shaped.}
\label{fig6}
\end{figure}

\begin{figure}[t]
\centering
\includegraphics[scale=1.1]{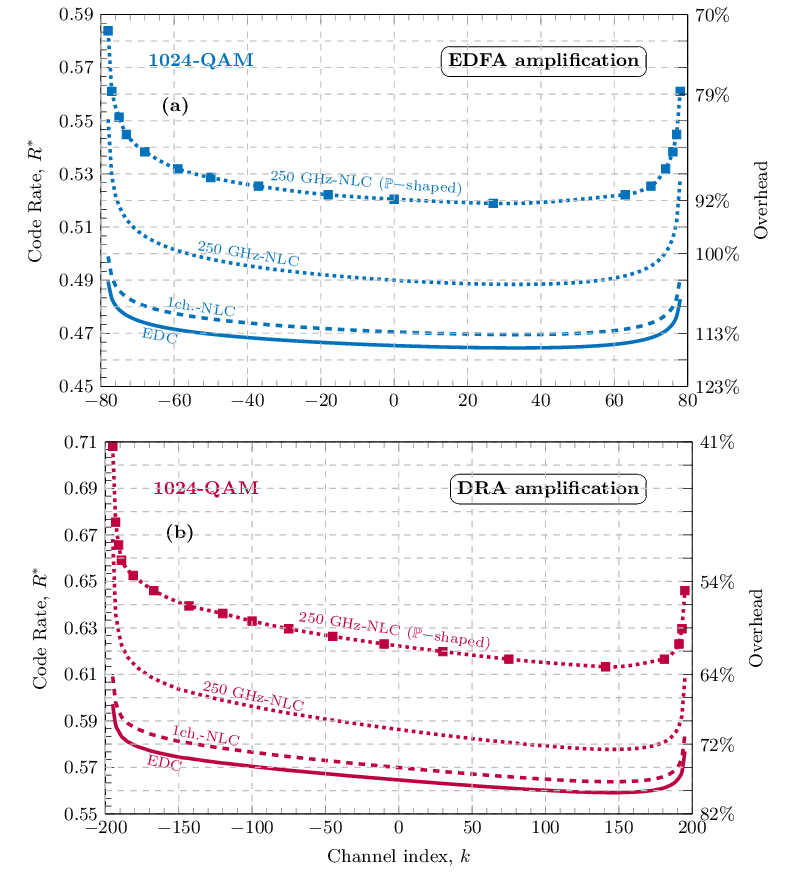}
\caption{Code rates (and overheads) for each individual channels for DP-1024QAM optical communication systems. (a) EDFA system (b) Raman amplified system. $\mathbb{P}-$shaped: probabilistically shaped.}
\label{fig7}
\end{figure}

The SNR at optimum launched power in each individual channel has been evaluated for both EDFA and Raman amplified optical communication systems as shown in Fig.~\ref{fig1}. It can be observed that, compared to the reported analyses (the green dashed curves) \cite{Nikshev2016, Semrau2017}, the impact of $\beta_{3}$ is reflected in the tilt in the nonlinear distortion spectrum due to the low frequency components exhibiting a greater accumulated dispersion, which ultimately results in a lower influence of nonlinear interference. This gives rise to a different performance for each channel in the WDM transmission system, which is no longer symmetric about the central channel.

The AIRs and the achievable SE (right-hand side) for each individual channel have been investigated for the modulation formats of DP-64QAM, DP-256QAM and DP-1024QAM, as shown in Fig.~\ref{fig2}, Fig.~\ref{fig3} and Fig.~\ref{fig4}. It can be seen that the AIR of each WDM channel is not symmetric about the central channel anymore as a consequence of the tilt in the nonlinear interference spectrum. This is fundamentally different from reported results in \cite{Poggiolini2014,Bosco2011,Curri2013,Nikshev2016, Semrau2017}, where the performance of WDM channels behaved symmetrically about the central channel. For both EDFA and Raman amplified transmission systems, the impact of third-order dispersion becomes more significant for the systems using higher-order modulation formats and higher-bandwidth NLC, because the performance of such systems are more sensitive to channel distortions.

Reciprocally, the ideal code rates $R^{\ast}$ and coding overheads ($\mathrm{OH} \triangleq \left(\left(1/R^{\ast}\right)-1\right)\cdot100\%$) for each individual channel were analyzed for different modulation formats in both EDFA and Raman amplified systems. Figure~\ref{fig5}, Fig.~\ref{fig6} and Fig.~\ref{fig7} illustrate that the ideal FEC code rate (or coding overhead) in each WDM channel is entirely asymmetric about the central channel due to the effect of the third-order dispersion. For the scheme of Raman amplified DP-1024QAM communication system with 250~GHz NLC in Fig.~\ref{fig7}(b), the difference between the code rates of the two outer channels is equal to 0.06, which is of great importance for FEC coding design.

It is found that, for all considered modulation formats, the ideal code rates behave significantly asymmetric with respect to the central channel due to the impact of third-order dispersion. This effect is observed to be more significant for higher-order modulation formats. This clearly indicates that the code rates in the forward error correction schemes have to be optimized asymmetrically in each individual channel in order to maximize the overall throughput.

The AIRs, achievable SEs, code rates and FEC overheads for both EDFA and DRA amplified transmission schemes with 250~GHz NLC are also illustrated, as shown in from Fig. \ref{fig2} to Fig. \ref{fig7}, when the discrete Maxwell-Boltzmann probabilistic shaping is applied. It can be found that, in all scenarios, the gains in AIRs, achievable SEs, code rates and FEC overheads increase with the order of signal modulation formats. Meanwhile, the asymmetric behaviors (in terms of AIRs, achievable SEs, code rates and FEC overheads) are very significant as well in such probabilistically shaped transmission schemes, when the third-order dispersion is taken into consideration.

\section{Numerical Simulations}

Wideband numerical simulations have also been implemented to verify the impact of the third-order dispersion (chromatic dispersion slope). The investigated scenario was a Nyquist-spaced WDM optical transmission system, using DP-64QAM, DP-256QAM and DP-1024QAM with parameters shown in Table \ref{tab2}. The transmitted symbol sequences in each channel and polarization are independent and randomly generated. The signal propagation was simulated in both standard single-mode optical fiber (SSMF) and non-zero dispersion-shifted fiber (NZDSF) using the split-step Fourier method for the Manakov equation, where a logarithmic step-size distribution was adopted for each fiber span \cite{bosco2000}. EDFAs were used to compensate for the fiber attenuation in each span. At the receiver, the signal was mixed with an ideal free-running local oscillator (LO) to ensure ideal coherent detection of the optical signal. Here only the EDC was performed using an ideal frequency domain filter. The SNR is estimated based on the received symbols in each individual channel. All numerical simulations are implemented with a digital resolution of 162 sample/symbol to guarantee the accuracy. The phase noise and the frequency offset from the transmitter and LO lasers, and the differential group delay (DGD) between the two polarization states in the fiber are neglected. The simulated SNRs of 81-channel transmission systems with and without the dispersion slope have been shown in Fig.~\ref{fig8}. It can be found that numerical simulations basically show good agreements in terms of the SNR behaviors compared to the analytical model. The small discrepancies (less than 0.3~dB for all channels) originate from the non-Gaussian input signal modulation and are in consistent with the reported results \cite{semrau2019}.

\begin{table}[t]
\caption{System and model parameters}
\begin{center}
\begin{tabular}{l|cc|c}
\toprule
Parameters & SSMF & NZDSF & Units  \\
\midrule
\midrule
Carrier wavelength     & \multicolumn{2}{c|}{1550}   & nm \\
Fiber attenuation      & 0.2 & 0.19 & dB~km\textsuperscript{-1}  \\ 
Fiber dispersion       & 17 & 4.5 & ps~nm\textsuperscript{-1}~km\textsuperscript{-1}\\ 
Fiber dispersion slope & 0.067 & 0.05  & ps~nm\textsuperscript{-2}~km\textsuperscript{-1} \\ 
Fiber nonlinearity     & 1.2 & 1.3 & W\textsuperscript{-1}~km\textsuperscript{-1} \\ 
Gain slope             & 0.028 & 0.031  &  W\textsuperscript{-1}~km\textsuperscript{-1}~THz\textsuperscript{-1}  \\
Launched power per channel         
                       &  -- 2.76 & -- 5.10 & dBm \\
Symbol rate            & \multicolumn{2}{c|}{32}     & GBd \\ 
Number of channels     & \multicolumn{2}{c|}{81}     & -- \\ 
Roll-off factor        & \multicolumn{2}{c|}{0.01}   & \%  \\ 
EDFA~NF                & \multicolumn{2}{c|}{4.5}    & dB  \\ 
Number of symbols      & \multicolumn{2}{c|}{$2^{18}$} & -- \\
Transmission distance  & \multicolumn{2}{c|}{15$\times$80}   & km \\ 
Number of steps per span   &  \multicolumn{2}{c|}{$8\times10^{6}$}  & [log-scale] \\
\bottomrule
\end{tabular}
\label{tab2}
\end{center}\vspace*{-18pt}
\end{table}

\section{Discussions}

\begin{figure}[t]
\centering
\includegraphics[scale=1.1]{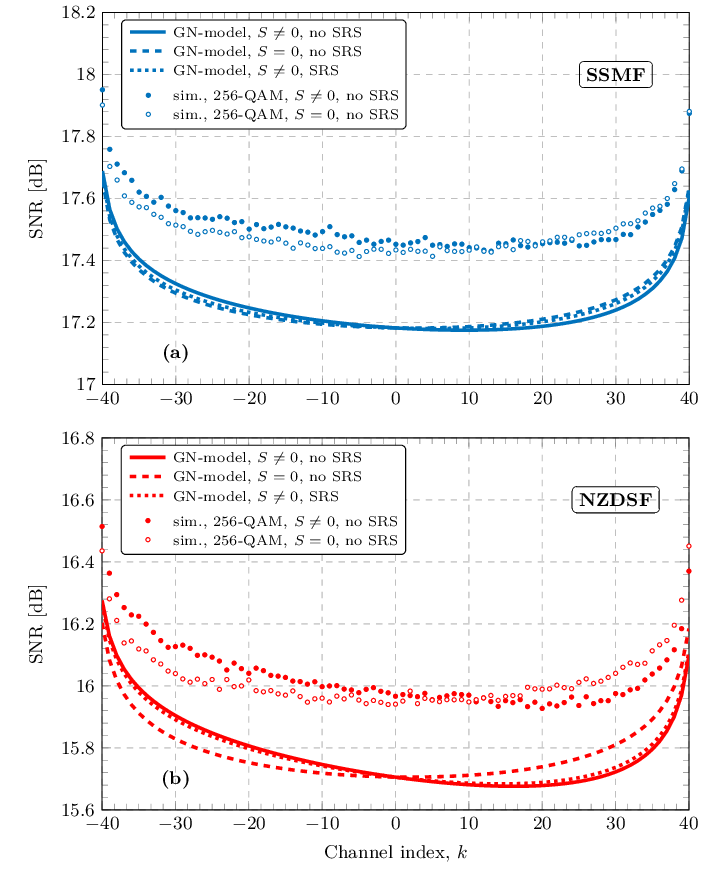}
\caption{The comparisons of DP-256QAM signalling split-step simulations with analytical modelling in terms of SNR against WDM chanel index performance after $15\times80$~km transmission distance for a fixed centre channel \emph{optimum} launched power considering two types of optical fibres (a): SSMF (b): NZDSF.}
\label{fig8}
\end{figure}

\begin{figure}[t]
\centering
\includegraphics[scale=0.75]{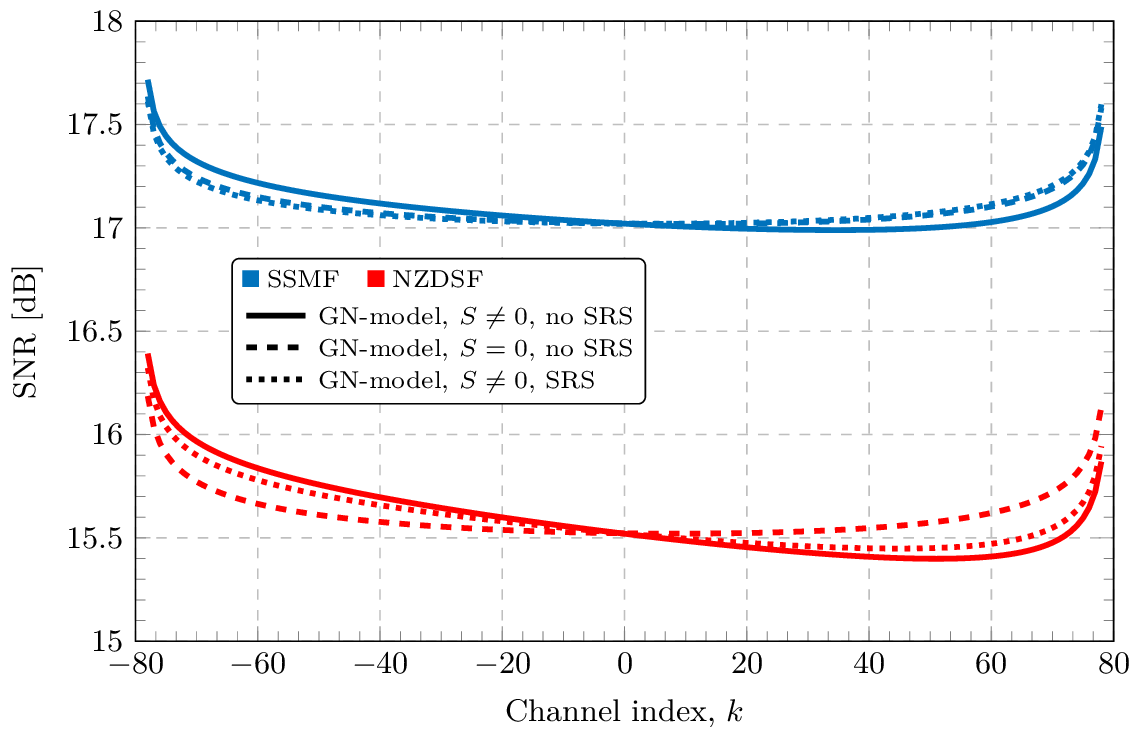}
\caption{Theoretical SNR versus channel index including the impact of the inter-channel SRS effect. The curves are calculated for $15\times80$~km $C-$band lumped amplified transmission system considering SSMF and NZDSF.}
\label{fig9}
\end{figure}

The impact of the inter-channel stimulated Raman scattering (SRS) effect has also been investigated and discussed based on the developed SRS model \cite{semrau2019,roberts2017,cantono2018}. Without the loss of generality, we investigated the SSMF and the DZDSF transmission schemes with a lumped amplification scenario. The results of the 81-channel transmission are shown in Fig.~\ref{fig8} and the results of the 161-channel ($C-$band) transmission are shown in Fig.~\ref{fig9}. Here the optical signal powers were optimised considering the third-order dispersion only. It can been seen in Fig.~\ref{fig8} that the inter-channel SRS gives a negligible impact on our considered transmission scheme for both SMF and NZDSF when the third-order dispersion is applied. The transmission bandwidth is further increased to $C-$band 157-channel to study the impact of SRS. It is found in Fig.~\ref{fig9} that the impact of SRS becomes slightly larger but still quite marginal, when the transmission bandwidth is smaller than 5~Hz and the optical signal powers are optimised considering the third-order dispersion only.

\section{Conclusion}

In summary, the impact of Kerr fiber nonlinearities in $C-$band ($\sim$5~THz) EDFA and $C+L-$band ($\sim$12.5~THz) Raman-amplified optical communication systems are investigated considering the effect of third-order chromatic dispersion. The achievable information rate and the FEC code rates in each individual transmission channel have been evaluated for DP-64QAM, DP-256QAM and DP-1024QAM modulation formats, when the probabilistic shaping technique has been applied. It is found that, for all considered modulation formats, the AIR and the ideal FEC code rates are asymmetric about the central channel due to the impact of the third-order dispersion. This provides us a new insight that the FEC schemes have to be optimized asymmetrically, almost on a per-channel basis, in order to maximize the overall throughput of the transmission system.

\ifCLASSOPTIONcaptionsoff
  \newpage
\fi

\bibliographystyle{IEEEtran}

\end{document}